\documentclass[twocolumn,letterpaper,10pt]{article} 
\usepackage{iasted} 
\usepackage{times} 
\usepackage[dvips]{graphicx} 

\usepackage{zedter}
\usepackage{z-eves}

\usepackage{diagrams}

\def\apll{ ]\!\!|\!\![~} 
\usepackage{boxedminipage}

\begin{document} 
 
\date{}

\title{Tool-assisted Multi-facet Analysis of Formal Specifications
(using {Atelier-B} and {ProB})} 


\author{
Christian Attiogb\'e - Christian.Attiogbe@univ-nantes.fr\\
LINA  - FRE CNRS 2729 \\
Pre-publication version @ SE'06}

 
\maketitle 
 
\thispagestyle{empty} 
 
\noindent 
{\bf\normalsize ABSTRACT}\newline 
{Tool-assisted analysis of software systems and convenient guides to practise the formal methods are still motivating challenges.
This paper addresses these challenges and shows using a case study that one can increase the quality of the software by analysing from multiple aspects a formal specification.
The B method and the Atelier-B tool are used for formal specifications, for safety property analysis and for refinements. The ProB tool is used to supplement the study with model checking; it helps to discover errors and therefore to improve the former specifications. } 
\vspace{2ex} 
    
\noindent 
{\bf\normalsize KEY WORDS}\newline 
{Formal Analysis, B Method, Theorem Proving, Model Checking.} 
 
\section{Introduction} 
The mechanization of software analysis  and software development process contributes in ensuring the quality of the developed software. In this paper we show  that one can increase the software analysis by combining several tools during the formal analysis process.\\
Our main motivation is the practical combination of development and formal analysis tools via their effective use on the projects at hand. 
Indeed a system in development may be analysed according to  various relevant aspects.
The analysed aspects are often safety properties and liveness properties. 
Theorem proving technique is used for consistency, refinement and safety property checking. Model checking is used for liveness property checking.
Very often these techniques require some specific input languages or models. 
As far as practical frameworks are concerned,  there is not a straightforward  interaction between the tools from both techniques; the users have to deal with different input specifications according to the targeted tool.
This is a flaw for their  complementarity on real developments; it may conduct to misleading results if there is no insurance that the starting models are the same ones. Moreover it is a burden for the developers. 

Multi-facet analysis consists in applying several appropriated analysis techniques and tools to the same formal model so as to study various aspects of it. This should be achieved  during a development process going from formal specification to concrete code via analysis.

We show using a small case study how such multi-facet analysis can be dealt with; we use for this purpose Atelier-B (a theorem prover and refinement checker) and ProB (an animator and model checker). Here B specifications serve as a reference model for the performed analyses.

The B method \cite{Abr96} ranges in the family of development methods founded on refinement techniques. Theorem proving is used to discharge refinement correctness proof obligations as well as safety property proofs.
Further developments of the B method, known as Event B \cite{Abr96a,AbrialMussat98,AbrialDSM2002}, enable one to develop more generally discrete event systems. 
The Event B approach is very close to Action Systems \cite{BaKu83,BuMo95}. 
The B method is equipped with tools (Atelier-B\footnote{ClearSy, France, www.atelierb.societe.com}, B-Toolkit\footnote{B-Core, UK, www.b-core.com} are the industrial ones) which help the users to manage their B projects. 
However, these tools mainly deal with theorem proving techniques which are time consuming especially for large size projects. 
Multi-facet analysis may come into play in such projects; 
it may help to faster the analysis and to cover other aspects not covered by theorem provers. 
However, the choice of the  tools to be used for the multi-facet analysis is very important in order to reduce the development cost.
In the current case, the  ProB tool \cite{LeuschelButler:FME03,LeuschelTurner:ZB05} seems well adapted to supplement Atelier-B.    

The contribution of this article is many-fold:
\textit{i)} the case study reveals some interesting features of B specifications: for example, dealing with the dynamic generation of processes (which can interact with the already existing ones) in the specification;
\textit{ii)} the complementarity of the used tools increases the quality of the analysis;
\textit{iii)} a methodological guide to multi-facet analysis.

The paper is structured as follows. 
In  Section 2 the multi-facet analysis approach is presented. Section 3 presents the case study. In Section 4 we give an overview of Event B. The Section 5 is devoted to the ProB tool. In  Section 6 we present the complete formal analysis experiment.  
Section 7 is devoted to learned lessons and finally in Section 8  we give some concluding remarks.
\section{Multi-facet Analysis}
It is now established that formal specifications and their formal analysis until code generation are good practices of software engineering.
 However, the use of tools should accompany these practices so as to get some confidence. 
Very often the tools in place focus  only on specific analysis; that means one deals with a given facet of the considered system.\\ 
The motivation for the multi-facet analysis is the seek of a wide coverage of the various aspects that can be analysed when developing a large software. 

In \cite{AttiogbeSOFSEM05} we propose an approach for multi-facet analysis where we show that an \textit{abstract reference model} may be the basis of forthcoming specific models and the basis of various analysis tools, even if their input languages are different.
The proposed analysis approach consists to build a general reference model of the system at hand; to derive systematically the specific models of each involved analysis technique from the reference model; to perform the analysis directly on the derived specific models  or to extend them and then to perform the analysis on the extended specific models.
The result of each part of the analysis may help to tune the reference model.

A reference model (built for an entire system) may be used for both theorem proving  and model checking aspects. Typically a reference model is made of a static part and a dynamic one.
According to a system to be studied, the static part describes  the input, the output,  the state variables and the appropriate type information. It is basically an abstract state model from which isomorphic structures are derived to match with other specific formalisms.
The dynamic part is made of several operations which are viewed as a transition relation on the state space described with the variables of the static part. 

In the current study the used reference model is a B specification. It already has a static part and a dynamic one. According to the chosen tools which use B specifications as input, we have not derive specific models; we perform directly the analyses on the same B specifications. The feedbacks from the used tools help to improve the reference model.
\section{Brief Presentation of the Case Study}
The current case study is a highly practical problem: the Readers/Writers well-known problem.
It is of some interest for distributed software which can be encountered within Internet services for example to deal with the accesses to resources. 

The Readers/Writers problem is a classical exclusion and synchronisation problem whose solutions are used in many application area such as database accesses by processes and operating systems. Two kind of \textit{processes} are considered: the writer processes and the reader processes.
The writing (by the writer processes) is exclusive but several readers can read together.
For example the file system manager of an operating system implements a file-system locks logic in order to prevent for inconsistencies on the files which are shared resources for the concurrent processes.  An exclusion policy and a synchronisation mechanism are necessary because an inconsistency may arise if some processes (the readers) read a resource when other processes (the writers) write the resource.

\noindent
\textit{Analysis} There are several policies for this resource management logic. 
A simple one  allows an arbitrary number of readers or alternatively only one writer to access a resource at a given time. 
This simple policy ensures the integrity of the resource but is not satisfactory; it gives readers preference over writers. Consider for example the case where there are several readers which want to access the resource; if at any given time there is one reader which is reading, the readers take the priority over the writers.
The latter ones may be starved (undesired property).

One of the solution often used to deal with this situation is to bound the number of consecutive readers which access the resource.
 Therefore, the writers may catch the priority when the number of consecutive readers is reached. This solution is rather fair and does not starve the writers. 

In order to develop software systems which integrate such access policies, it is worthwhile to study them carefully before their implementation and deployment. 
A formal study of the desired software may help to discover faults and to correct them. 
It may also help to develop the right software. 
That is dealt with in the remainder of the paper.

Note that behind the simple appearance of this case study there are some interesting formal specification features. 
Not all specification formalisms can be used to specify such a simple but real system: within an operating system for example, the processes are created at any time, they run and they terminate. 
The number of the processes is not known in advance;  
their behaviours are not completely known; but they can interact with the already existing processes.
 
Most of the specification formalisms (for example process algebra) does not deal with the composition of undefined number of processes; they require that not only  the process behaviours but their number be defined before their composition.
Moreover there is not a simple way to deal with dynamic process creation. For example, a system will be studied with a given number of readers and writers. Consequently some aspects which are related to dynamic process creation are not well studied.

Therefore the use of the Event B approach here is relevant. We can take into account these specific aspects.  
   
Before the illustration of the multi-facet analysis on this problem we give a brief presentation of the Event B method which is used for the specification.
\section{Overview of Event B}
\label{section:overview_EB}
Within the  Event B framework,  asynchronous systems may be developed and structured using \textit{abstract systems}. 
\textit{Abstract systems} are the basic  structures of the so-called \textit{event-driven} B, and they replace the \textit{abstract machines} which are the basic structures of the earlier \textit{operation-driven} approach of the B method. \\ 
An abstract system describes a mathematical model of an asynchronous system behaviour\footnote{A system behaviour is the set of its possible transitions from state to state beginning from an initial state.}; it is  mainly made of a state description (constants, properties, variables and invariant) and several \textit{event} descriptions.
Abstract systems are comparable to Action Systems \cite{BaKu83}; they describe a non-deterministic evolution of a system through guarded actions. 
Dynamic constraints can be expressed within abstract systems to specify various liveness properties \cite{AbrialMussat98}. 
The state of an abstract system is described by variables and constants linked by an invariant. Variables and constants represent data of the system being formalized. Abstract systems may be refined like abstract machines. 

\noindent
\textbf{Data and events of an abstract system} Unlike an abstract machine, an abstract system does not encapsulate the data protected by the operations; it rather adopts a global vision and contains global data of an entire model, be it distributed or not. The data are described with sets, predicates, relations and functions.\\
Within the B approach, an event is considered as the observation of a system transition. Events are  spontaneous and show the way a system evolves. \\
An event has a \textit{guard} and an \textit{action}. It may occur or may be observed only when  its guard holds. 
An event is $enabled$ if its guard holds otherwise the event is $disabled$. 
The action of an event describes using generalized substitutions \cite{Abr96} the way in which the state of the system evolves when this event occurs.
Several events can have their guards held simultaneously; in this case, only one of them  occurs. The system makes internally a non-deterministic choice. If no guard is true the abstract system is blocking (deadlock).

\begin{figure}[htp]
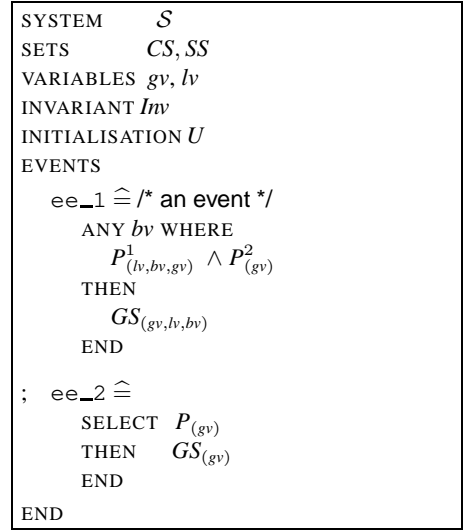

\noindent
{\small
\begin{center}
\begin{boxedminipage}{6cm}
\textsc{system}  ~~~~~ ~~${\mathcal S}$\\
\textsc{sets} ~~~~~$~ ~ ~ ~ ~ ~ ~ ~ ~ ~CS, SS$\\
\textsc{variables } $gv$, $lv$\\
\textsc{invariant} $Inv$\\
\textsc{initialisation} $U$\\
\textsc{events}
\vspace{-0.2cm}
\begin{tabbing}
\hspace{0.4cm}\=\hspace{0.4cm}\=\hspace{0.4cm}\=\hspace{0.4cm}\kill
\>\texttt{ee\_1} $\defs$ \textsf{/* an event */}\\
\>\>\textsc{any} $bv$ \textsc{where}\\
\>\>\> $P^{1}_{(lv, bv, gv)}~ \land P^{2}_{(gv)}$\\
\>\>\textsc{then} \\
\>\>\>$GS_{(gv, lv, bv)}$\\
\>\>\textsc{end}
\end{tabbing}
\vspace{-0.4cm}
\begin{tabbing}
\hspace{0.4cm}\=\hspace{0.4cm}\=\hspace{0.4cm}\=\hspace{0.4cm}\kill
;\>\texttt{ee\_2} $\defs$  \\
\>\>\textsc{select}
~~$P_{(gv)}$ \\
\>\>\textsc{then} 
~~~~     $GS_{(gv)}$\\
\>\>\textsc{end}
\end{tabbing}

\vspace{-0.3cm}
\textsc{end}
\end{boxedminipage}
\end{center}}
\caption{Shape of an abstract system}
\label{figure:canonicS1andS2}
\end{figure}

The shape of an abstract system is given  in Figure \ref{figure:canonicS1andS2}.
An event has one of the general forms given with \texttt{ee\_1} or \texttt{ee\_2} (see Fig. \ref{figure:canonicS1andS2}) where $bv$ denotes the bound variables of the event (their are bound to the ANY) ; $gv$ denotes the global  variables of the abstract system containing the event. 
$P_{(bv, gv)}$ is a predicate over the variables $bv$ and $gv$; 
$GS_{(bv, gv)}$ is a generalized substitution (or the parallel composition of generalized substitutions) which models the event action using $bv$ and $gv$.
The SELECT form (\texttt{ee\_2}) is  a particular case of the ANY form (\texttt{ee\_1}). 
The guard of an event with the SELECT form is $P_{(gv)}$.
The guard of an event with the ANY form is $\exists(bv).P_{(bv,gv)}$.

\noindent
\textbf{Semantics and consistency} 
An Event B model is described by an abstract system. 
The semantics of an abstract lies in its invariant and is guaranteed  by proof obligations. 
The consistency of the model is also established by proof obligations.  
For the consistency of an Event B model, the initialization should establish the invariant and each event of the given abstract system should preserve the invariant of the model; one must prove these obligations. In the case of an event with the ANY form, the proof obligation is:
 {\small $$Inv_{(gv)} \land P_{(bv,gv)} \implies [GS_{(bv,gv)}]Inv_{(gv)}$$} where $Inv_{(gv)}$ stands for the invariant of the abstract system. 
The notation $[Subst]Predicate$ denotes a predicate transformer. 
It expresses that the substitution $Subst$ establishes $Predicate$.

\noindent
\textbf{Composition of Abstract Systems} 
We  introduced in \cite{AttiogRR0208,AttiogbeQSIC04,Attiogbe:ZB05} a parallel composition operator to supplement the classical top-down approach of the Event B. This parallel composition allows communication through global variable shared by the abstract (sub)systems. 
This permits a bottom-up approach for the Event B. 
For the composition an abstract system  ${\mathcal S_i}$ is described by a signature and a body: ${\mathcal S_i} = \langle signature({\mathcal S_i}), body({\mathcal S_i}) \rangle$. 
The body $body({\mathcal S_i})$ is made of the variables ($V_i$), the invariant ($Inv_i$), the initialization ($U_i$) and the set of events ($E_i$) of the abstract system. 
Thus an abstract system  ${\mathcal S_i}$ is described with $\langle \Sigma_i, \langle V_i, Inv_i, U_i, E_i \rangle \rangle$.

As far as the composition is concerned, the invariant $Inv_i$ of ${\mathcal S_i}$ is rewritten with local and common state variables of  ${\mathcal S_i}$ as: ${I_{(gv)}}_i \land {J_{(lv)}}_i \land {K_{(gv, lv)}}_i$. \\
${I_{(gv)}}_i$ is the common part of the invariant shared by the considered abstract systems.
${J_{(lv)}}_i$ deals with the local properties,  
${K_{(gv, lv)}}_i$ relates the local variables and the global ones and it expresses the associated properties.  

\noindent
The parallel composition of two abstract systems   ${\mathcal S_1} = \langle \Sigma_1, \langle V_1, Inv_1, U_1, E_1 \rangle \rangle$ and  ${\mathcal S_2} = \langle \Sigma_2, \langle V_2, Inv_2, U_2, E_2\rangle \rangle$ is defined as:\\

{\small 
\begin{tabular}{c}
$\langle \Sigma_1 \cup \Sigma_2, \langle V_1 \cup V_2, Inv_1 \wedge Inv_2, U_1 \| U_2, E_1 \biguplus E_2 \rangle \rangle$
\end{tabular}
$~$\\
}
It is noted ${\mathcal S_1}$ $\apll$ ${\mathcal S_2}$.
\noindent
This parallel composition is commutative and associative; therefore it is also defined for $n$ systems. More details on the parallel composition can be found in \cite{AttiogRR0208,Attiogbe:ZB05}.

The B method is supported by some theorem provers which are industrial tools (Atelier-B and B-Toolkit). The Event B extension has not dedicated tools but the specifications are translated into classical B and the standard B tools are used.
\section{Overview of the ProB Tool}
The ProB tool \cite{LeuschelButler:FME03,LeuschelTurner:ZB05} is an animator and a model checker for  B specifications. It provides functionalities to display graphical view of automata.
It supports automated consistency checking of B specifications. 
Here a B specification is an abstract machine (or a refinement) with its state space, its initialization  and its operations. The consistency checking is performed on all the reachable states of the machine. The ProB tool also provides a constraint-based checking; with this approach ProB does not explore the state space from the initializations, it checks whether applying one of the operation can result in an invariant violation independently from the initializations. 

The ProB offers many functionalities. The main ones are organized within three categories: \textit{Animation}, \textit{Verification} and \textit{Analysis}. Several functionalities are provided for each category: we list a few of them which are used in this paper.

As far as the  \textit{Animation} category is concerned, we have the following functionalities:
\begin{itemize}  
\item \textsf{Random Animation}: it starts from an initial state of the abstract machine and then, it selects in a random fashion one of the enabled operations, it computes the next state accordingly and proceeds the animation with one of the enabled operations with respect to the new state;  
\item \textsf{View/Reduced Visited States}: it displays a minimized graph of the visited states after an animation;
\item \textsf{View/Current State}: it displays the current state which is obtained after the animation.
\end{itemize}

In the  \textit{Verification} category, the following functionalities are available: 
\begin{itemize}  
\item \textsf{Temporal Model Checking}:  starting from a set of initialization states (initial nodes), it systematically explores the state space of the current B specification. From a given state (a node), a transition is built for each enabled operation and it ends at a computed state which is a new node or an already existing one. Each state  is treated in the same way;
\item \textsf{Refinement Checking}:  the principle here is based on trace checking. All running traces of the refinement should be traces of the initial specification;
\item \textsf{Constraint Based Checking}: it checks for invariant violation when applying operation independently from initialization states.  
\end{itemize}

As far as the  \textit{Analysis} category is concerned, we have the following functionalities:  
\begin{itemize}  
\item \textsf{Compute Coverage}: the state space (the nodes) and the transitions of the current specification are checked, some statistics are given on deadlocked states, live states\footnote{those already computed}, covered and uncovered operations; 
\item \textsf{Analyse Invariant}: it checks if some parts of the current specification invariant are true or false; 
\item \textsf{Analyse Properties}: if the current specification has a property clause, then it is checked.
\end{itemize}
Note that if a B prover has been used to perform consistency proof, the invariant should not be violated; the B consistency proof consists in checking that the initialization of an abstract machine establishes the invariant and that all the operations preserve the invariant. Therefore in the case where the consistency is not completely achieved ProB can help to discover the faults.  
\section{Formal Analysis of the Readers/Writers}
We present here the main steps of our experiment: abstract specification with Event B, refinement, property analysis. 

\subsection{B Specification: a Reference Model}
The Event B approach allows us to specify the asynchronous behaviour of processes. Moreover we can specify the interaction between unbounded number of such asynchronous processes using guarded nondeterministic generalized substitution (the ANY structure). 

A stepwise approach is used. We follow the structure of the real system by considering two kind of processes: the readers and the writers.  
The remainder of the study follows this structure.

\noindent
\textbf{First step: Abstract Specification}
We specify two B abstract systems: the first one is  named $Readers$, the second is named $Writers$. They use the same sets READER and WRITER and their specifications are quite symmetric. The sets  READER and WRITER are the abstraction of considered processes.
The shape of the \textit{Readers} abstract system is depicted in the Figure \ref{figure:readers}. The \textit{Writers} is similar.

\begin{figure}
{\small
\begin{center}
\begin{boxedminipage}{8cm}
\textsc{system}  ~~~~~~~$Readers$\\
\textsc{sets} ~~~~~$~ ~ ~ ~ ~ ~ ~ ~ ~ ~READER; WRITER$\\
\textsc{variables } \\
 $~$ \hfill $readers, waitingReaders,activeReaders,activeWriter$\\
\textsc{invariant} $~~~\cdots$\\
\textsc{initialisation} $~~~\cdots$\\
\textsc{events}
\vspace{-0.2cm}
\begin{tabbing}
\hspace{0.4cm}\=\hspace{2.4cm}\=\hspace{0.4cm}\=\hspace{0.4cm}\kill
\>\texttt{want2read} \> $\defs$ $~~~\cdots$ \\
;\>\texttt{reading} \> $\defs$  $~~~\cdots$\\
;\>\texttt{endReading} \> $\defs$  $~~~\cdots$\\
;\>\texttt{newReader} \> $\defs$  $~~~\cdots$\\
\end{tabbing}
\vspace{-0.5cm}
\textsc{end}
\end{boxedminipage}
\end{center}}
\caption{B specification of the readers}
\label{figure:readers}
\end{figure}

The variable names are self-explanatory; the variable $readers$ describes the current readers (abstraction of the reader processes). 
The  invariant of the $Readers$ abstract system is as follows:
{\small
\begin{center}
\begin{boxedminipage}{8cm}
\textsc{invariant}\\
\begin{tabular}{l}
$~~~readers \subseteq READER$ 
$\land	waitingReaders \subseteq READER$\\
$\land	activeReaders \subseteq READER$ \\
$\land	card(activeReaders) \ge 0$\\
$\land	readers \cap waitingReaders = \{\}$\\
$\land	activeReaders \cap waitingReaders = \{\}$\\
$\land	activeReaders \cap readers = \{\} $\\
$\land	activeWriter \subseteq WRITER $ 
$\land	card(activeWriter) \le 1$\\
$ \land	\lnot( (card(activeWriter) = 1) \land $\\
$~~~~~~~~~~~~~~(card(activeReaders) \ge 1))$ \\
\end{tabular}
\end{boxedminipage}
\end{center}}

\noindent
\textit{Exclusion property:}\\
The predicate $not( (card(activeWriter) = 1) \land (card(activeReaders) \ge 1))$ which appears in the invariant describes the exclusion property required for the readers and the writers. 
It states that we should not have active readers together with a writer. This is a correctness property included in the invariant. 

\noindent
The abstract system $Readers$ has the events {\small \texttt{want2read}}, {\small \texttt{reading}}, {\small \texttt{endReading}} and {\small \texttt{newReader}}. 
The event {\small \texttt{want2read}} occurs when a given readers $rr$ wants to read; it is then recorded in the variable $waitingReaders$ and should begin the reading as soon as possible according to the global system state. \\

{\small
\begin{center}
\begin{boxedminipage}{8cm}
\begin{tabbing}
\hspace{0.4cm}\=\hspace{0.4cm}\=\hspace{0.4cm}\=\hspace{0.4cm}\kill
\>\texttt{want2read} $\defs$ \\
\>\>\textsc{any} $rr$ \textsc{where}\\
\>\>\> $rr \in readers \land rr \notin waitingReaders$\\
\>\>	$\land$\>$ rr \notin activeReaders$\\
\>\>\textsc{then} \\
\>\>\>$ waitingReaders := waitingReaders \lor \{rr\}$\\
\>\>$\|$\> $readers := readers - \{rr\}$\\
\>\>\textsc{end}
\end{tabbing}
\end{boxedminipage}
\vspace{-0.3cm}
\end{center}
}

\vspace{0.2cm}
The event {\small \texttt{reading}} is specified as follows:
{\small
\begin{center}
\begin{boxedminipage}{8cm}
\begin{tabbing}
\hspace{0.4cm}\=\hspace{0.4cm}\=\hspace{0.4cm}\=\hspace{0.4cm}\kill
\>\texttt{reading} $\defs$ \\
\>\>\textsc{any} $rr$ \textsc{where}\\
\>\>\> $rr \in waitingReaders \land activeWriter = \{\}$\\
\>\>\textsc{then} \\
\>\>\>$   activeReaders := activeReaders \cup \{rr\}$\\
\>\>$\|$\> $waitingReaders := waitingReaders - \{rr\}$\\
\>\>\textsc{end}
\end{tabbing}
\end{boxedminipage}
\vspace{-0.3cm}
\end{center}
}

\medskip
The guard of the event {\small \texttt{reading}} states that: any reader $rr$  which is waiting for reading, and such that there is no active writer, can start reading.
Note that we have not specify the behaviour of only one specific reader process, but  that of any number of readers. It is achieved trough  the use of non-determinism to specify the events of the abstract system.

The  invariant of the second abstract system ($Writers$) is quite similar to that of $Readers$ but it has the constraint of exclusion between writers: $card(activeWriter) \le 1$.

The $Writers$ abstract system  has the same shape as the $Readers$; it has the events {\small \texttt{want2write}}, {\small \texttt{writing}}, {\small \texttt{endWriting}} and {\small \texttt{newWriter}}. 
The event {\small \texttt{writing}} is as follows:
{\small
\begin{center}
\begin{boxedminipage}{8cm}
\begin{tabbing}
\hspace{0.4cm}\=\hspace{0.4cm}\=\hspace{0.4cm}\=\hspace{0.4cm}\kill
\>\texttt{writing} $\defs$ \\
\>\>\textsc{any} $ww$ \textsc{where}\\
\>\>\> $ww \in waitingWriters \land activeReaders = \{\}$\\
\>\>\textsc{then} \\
\>\>\>$   activeWriter :=  \{ww\}$\\
\>\>$\|$\>$ waitingWriters := waitingWriters - \{ww\}$\\
\>\>\textsc{end}
\end{tabbing}
\end{boxedminipage}
\end{center}
}

Thereafter we parallely combine these abstract systems to get the global interacting system named $readWrite$.\\

\begin{boxedminipage}{7.5cm}
\centerline{$readWrite \defs Readers \apll Writers$}
\end{boxedminipage}

\medskip
The resulting abstract system  conjoins the invariants of the subsystems; its events consist of the merging of the events from $Readers$ and $Writers$.

Note that in this first step of the study we consider the simple policy (an arbitrary number of readers or alternatively only one writer) to define a first working solution; thereafter we refine this solution to take into account the fairness property.

\noindent
\textbf{Second step: Refinement}
We refine the specifications which result from the first step; we include a new variable $nbConsecutiveReaders$, a constant $maxConsecutiveR$ and a new property in the invariant. We also introduce two new events: {\small \texttt{leaveReader}} and  {\small \texttt{leaveWriter}}. The latter ones simulate the destruction of processes.
The included property is about fairness: not only the readers or the writers are allowed to run. This is dealt with by limiting (with $maxConsecutiveR$) the number of consecutive accesses for the readers. 

The events and their guards are rewritten and updated according to the new variables. Again we use a bottom-up stepwise approach: the abstract systems $Readers$ and $Writers$ are respectively  refined into  $ReadersR$ and $WritersR$. These latter ones are  parallely combined to get the global interacting system named $readWriteR$. In the refinement the variable $nbConsecutiveR$ is set to 0 when a writer get the access for writing.

\noindent
\begin{boxedminipage}{8cm}
\centerline{$readWriteR \defs ReadersR \apll WritersR$}
\end{boxedminipage}

\medskip
\noindent
The invariant of the global refined system is as follows:\\
{\small
\begin{boxedminipage}{8cm}
\textsc{invariant}\\
\begin{tabular}{l}
$ ~~~	writers \subseteq WRITER $ \\
$ \land	activeWriter \subseteq WRITER $ \\
$ \land	card(activeWriter) \le 1$ \\
$ \land	waitingWriters \subseteq WRITER$ \\
$ \land	writers \cap waitingWriters = \{\}$ \\
$ \land	activeWriter \cap waitingWriters = \{\}$ \\
$ \land	activeWriter \cap writers = \{\} $ \\
$ \land	nbActiveReaders \in NAT$ \\
$ \land	nbActiveReaders = card(activeReaders)$ \\
\textbf{$\land$	nbConsecutiveR $\in$ NAT} \\
\textbf{$\land$ nbConsecutiveR $\le$ maxConsecutiveR} \\
$ \land	readers \subseteq READER$ \\
$ \land	waitingReaders \subseteq READER$ \\
$ \land	activeReaders \subseteq READER$ \\
$ \land	card(activeReaders) \ge 0$ \\
$ \land	readers \cap waitingReaders = \{\}$ \\
$ \land	activeReaders \cap waitingReaders = \{\}$ \\
$ \land	activeReaders \cap readers = \{\}$ \\ 
$ \land	\lnot( (card(activeWriter) = 1) \land$\\
$~~~~~~~~~~~~~~~~(card(activeReaders) \ge 1))$ \\
\end{tabular}
\end{boxedminipage}
}
 
\begin{figure}
\centerline{\resizebox{5.5cm}{3cm}{\includegraphics{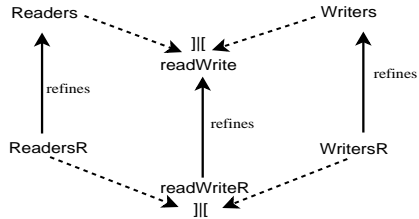}}}
\caption{Architecture of the B specification}
\label{figure:hierarchyR}
\end{figure}

\noindent
\textit{Architecture of the complete specification:}
The complete architecture of the system is depicted in the Figure \ref{figure:hierarchyR}.

\vspace{-0.3cm}
\subsection{Using Atelier-B} 
\noindent
Classical syntax and type checking are performed.\\
\noindent
\textbf{Consistency Checking and Safety Proof}
According to the B approach, all the events of the system should preserve the invariant. The proof obligations (POs) are generated for this purpose.  
POs generation is achieved and followed by consistency proof. 
The amounts of PO are displayed in the following tables.

{\small
\begin{tabular}{||c|c|c|c||}
\hline
\multicolumn{4}{|c|}{Abstract machines}\\
\hline
Components & {\small{$Readers$}} & {\small{$Writers$}} & {\small{$readWrite$}}\\
\hline
Nb. POs & 26  & 25 & 53 \\
\hline
\end{tabular}
}

Atelier-B  discharged automatically the major part of these POs. The rest is discharged using the interactive mode of the Atelier-B prover.\\
\noindent
\textbf{Refinement Proof}
For the refinement we have the following statistics.\\
{\small
\begin{tabular}{||c|c|c|c||}
\hline
\multicolumn{4}{|c|}{Refinements}\\
\hline
Components & {\small{$ReadersR$}} & {\small{$WritersR$}} & {\small{$readWriteR$}} \\ 
\hline
Nb. POs & 39 & 31 & 77\\ 
\hline
\end{tabular}
}

All the POs are discharged by combining the automatic mode and the interactive mode of the prover.

According to a classic formal analysis (without code generation) using the B prover, we may end our study there. 
But in the  current multi-facet analysis we follow the study by animating the model and model checking it using the ProB tool.
\vspace{-0.3cm}
\subsection{Using ProB} 
\begin{figure}
\centerline{\resizebox{8.5cm}{12cm}{\includegraphics{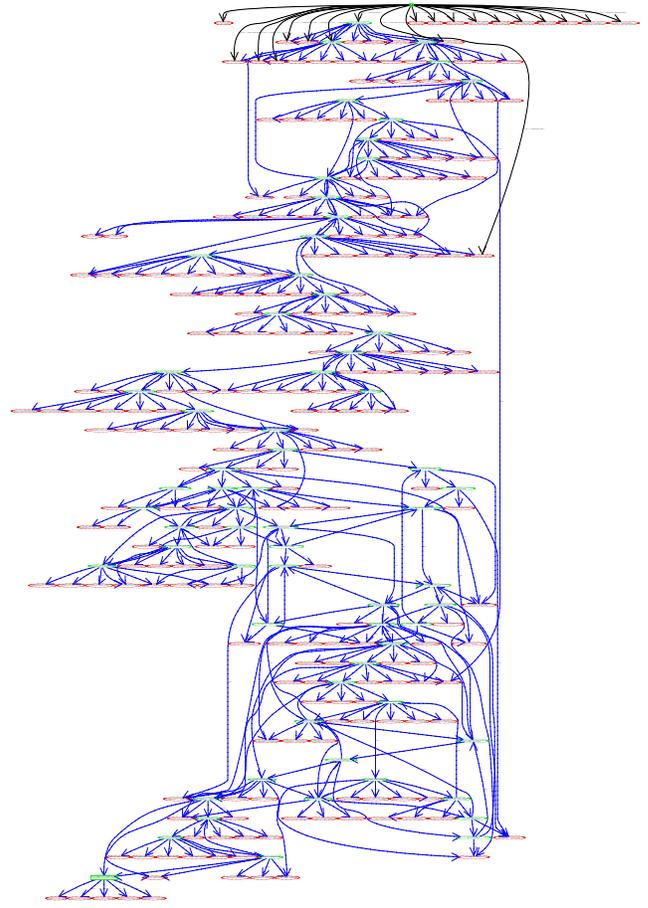}}}
\caption{Visited States with only 100 operations}
\label{figure:reducedViewFSM}
\end{figure}

With respect to the multi-facet analysis, the previous consistency checking and safety analysis based on theorem proving are increased using the ProB tool\footnote{We use here the version 1.1 of July 2004} to proceed with animation and model checking for liveness. We  describe here a part of the multiple analysis performed on the case study.
The Figure \ref{figure:reducedViewFSM} results from an experimentation using the $readWrite$ abstract system; it gives an idea of the explored states for a simulation with only 100 activated operations. The graphical views are unusable for higher parameters. 

\noindent
\textbf{Animation}
The model $readWriteR$ is animated using a random strategy (\textsf{Random Animation}) with a simulation parameterized  with 40 operation occurrences.

The result seems satisfactory; the computed state coverage is as follows: 
{\small
\begin{verbatim}
NODES
deadlocked          :0
invariant_violated  :0
explored_but_not_all_transitions_computed:1
live                :41
open                :129
total               :170
COVERED_OPERATIONS
setup_constants     :1
initialise_machine  :10
endWriting          :3
writing             :3
newReader           :13
leaveReader         :4
newWriter           :30
want2write          :15
leaveWriters        :15
want2read           :21
reading             :31
endReading          :32
UNCOVERED_OPERATIONS
\end{verbatim}
}
All the operations (the events) are covered. For each operation, the number behind the colon indicates how many times the operations occurred in the current simulation.
Note that there is no uncovered operation; also no deadlock is found ({\small \texttt{deadlocked :0}}).

\noindent
\textbf{Refinement Checking}
The entire state space of the abstract system $readWrite$ is explored; the resulting transition graph is stored and then the traces are compared with that of the refined system $readWriteR$. No errors was found.
 
\noindent
\textbf{Model Checking}
We follow the analysis by state exploration using the \textsf{Temporal Model Checking} functionality.\\
The \textsf{Compute Coverage} functionality shows the following status:
{\small
\begin{verbatim}
NODES
invariant_violated    :0
explored_but_not_all_transitions_computed:1
deadlocked            :1
live                  :1048
open                  :2174
total                 :3223
COVERED_OPERATIONS
...                   /* cut */
UNCOVERED_OPERATIONS
\end{verbatim}
}

\begin{figure}
\centerline{\resizebox{10.5cm}{1.5cm}{\includegraphics{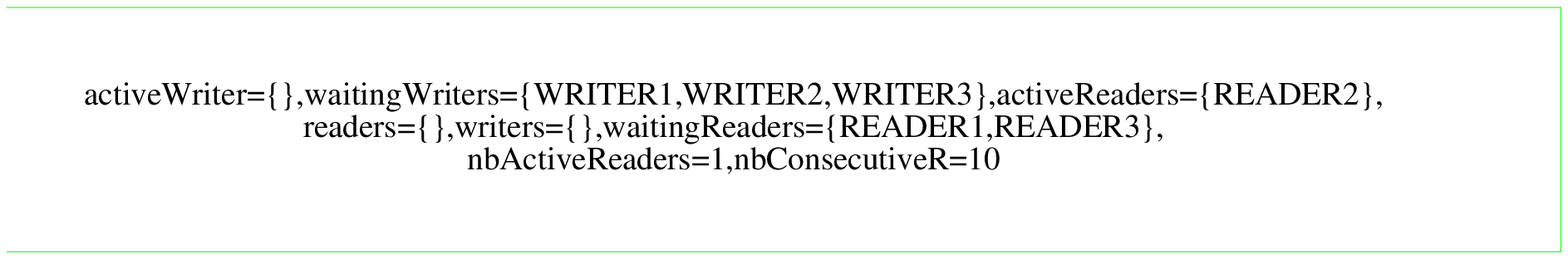}}}
\caption{Details of the deadlocked state}
\label{figure:deadlockState}
\end{figure}

A total of 3223 states is explored. It reveals that there is one deadlocked state ({\small \texttt{deadlocked :1}}).
We capture a view of the deadlocked state (see Fig. \ref{figure:deadlockState}).

The cause of the problem is identified when analysing this state: a reader process ({\small READER2}) is active and {\small $nbConsecutiveR=10$}; normally, the event {\small \texttt{endReading}} may occur but it does not. Thus we turn back to the specification which follows: 
{\small
\begin{boxedminipage}{8cm}
\begin{tabbing}
\hspace{0.4cm}\=\hspace{0.4cm}\=\hspace{0.4cm}\=\hspace{0.4cm}\kill
\>\texttt{endReading} $\defs$ \\
\>\>\textsc{any} $rr$ \textsc{where} /* an active reader finishes reading */ \\
\>\>\>$		rr : activeReaders $
      $\land	nbActiveReaders > 1$\\  
\>\>	\textsc{then}\\
\>\>\>$		activeReaders := activeReaders - \{rr\}$\\
\>\>$\|$\>$	readers := readers \cup \{rr\}$\\
\>\>$\|$\>$	nbActiveReaders := nbActiveReaders - 1$\\
\>\>	\textsc{end}
\end{tabbing}
\end{boxedminipage}
}
  
The problem with this event is now clear: the guard is wrong. We should have  $nbActiveReaders \ge  1$. 
Indeed the problem is that, when we have only one process which is reading, the guard of {\small \texttt{endReading}} is not true, and it cannot finish, hence the deadlock.
This is corrected in the specification and the problem is fixed.
This situation was not detected by the B prover. Indeed the consistency proof guarantees that the event does not violate the invariant. But the wrong guard does not violate the invariant.

This emphasizes the \textit{deadlock-freeness} proof obligation stated by Event B: 
\textit{"if no guard is true the abstract system is blocking (deadlock)"}. Therefore ProB (model checking) helps us to discover a proof obligation which was not discharged.  

\section{Learned Lessons}
We have presented one of the experimentations conducted using Atelier-B and ProB. The lessons we learnt from these experimentations (especially from the one presented here) are briefly summarized here.

Consistency proof using B theorem provers to discharge the generated proof obligations does not protect the user of specification faults. 
Animation helps to simulate some scenario of the developed system  but is not sufficient to ensure security.
Model checking by exploring the state space looking for deadlocked states (for example) may supplement the consistency checking performed by theorem proving; we showed that a wrong guard may not violate the invariant but it may lead to a deadlock. 
Model checking also help to speed up the task of theorem proving; if an error is discovered, the examination of the current state can be exploited to correct the specification.
More generally, a multi-facet analysis approach requires the use of a single reference model and several techniques and tools; therefore it helps to improve the quality of the specifications and consequently  that of the system which is being developed.
 
For large size system studies, it is more convenient to use the tools which operate on the same input formalisms; this helps in avoiding translation tasks and the propagation of feedback through various models.
In  previous work \cite{AttiogbeQSIC04} we used the \textsf{Spin} tool \cite{Holz97} to manage Event B specifications; even if the translation was systematic, it takes some intermediary steps to take account of the feedback. 

From the methodological point of view, this experiment shows a simple but efficient formal analysis process that can be reused to manage large systems:
analysis and specification of the system within B;
preliminary theorem proving tasks with the B prover;
model checking of the specification using ProB which is convenient for liveness properties.
In the case where errors are detected, the working steps are: display and examine carefully the current state with respect to the properties stated in the invariant; correct the specification accordingly and replay the process until no errors are detected.

\section{Discussion and Concluding Remarks}
We have used a multi-facet formal analysis approach to deal with the specification of software systems. The study is illustrated with the Readers/Writers problem; it involves composition, refinement, theorem proving and model checking so as to cover an entire development.
We specify the Readers, the Writers and then we combine them to have the global interacting system;  one important feature of this simple system is the dynamic process architecture (handling the composition of undefined number of processes).
The  Readers and  the Writers systems are refined to improve the model and to get some desired properties; the improvement concerns the introduction of some fairness and the introduction of new events.

Adopting the multi-facet approach forces the specifiers to  use several analysis techniques; we have used Atelier-B and the ProB tool starting from a single B specification considered as a reference abstract model. 
Using this approach, the complementarity of both tools is emphasized; more precisely we have improved the formal specification by supplementing the theorem proving (from Atelier-B) by model checking (from ProB). Independently from the classical invariant or properties violations that can be detected by model checking, we show that other specification faults (wrong guards of events) not detected by theorem proving are detected by the model checking. More precisely one proof obligation of Event B (\textit{deadlock-freeness}) is clearly handled by model checking. 

We have not explored all the capabilities of ProB, therefore future work will provide additional  assessment conclusions and more practical hints for formal analysis process.
In the context of medium  or large size development projects this multi-facet approach may be beneficially used to increase the software robustness.\\
Depending on the nature of the studied systems, the graphical visualization is more or less usable. A smaller one (see Fig. \ref{figure:reducedViewFSM}) from our experiment is quite unusable;
but related works \cite{LeuschelTurner:ZB05} are tackling this issue; the authors study the visualization of large state spaces using ProB. This will help for graph exploration.
\bibliographystyle{plain}

\end{document}